\documentclass[runningheads]{llncs}
\usepackage{graphicx}
\usepackage{enumitem}
\usepackage{amsmath,amssymb,amsfonts}
\usepackage{multirow}
\usepackage{booktabs}
\usepackage{array}
\usepackage{subcaption}
\usepackage{tikz}
\usepackage{xcolor}
\usepackage{algorithm}
\usepackage{algpseudocode}
\usepackage{float}
\usepackage{marvosym}
\usepackage{cite}

\begin{document}
\title{LADSG: Label-Anonymized Distillation and Similar Gradient Substitution for Label Privacy in Vertical Federated Learning}
\titlerunning{LADSG: Label Privacy in Vertical Federated Learning}

%
\author{
Zeyu Yan\inst{1}\textsuperscript{†} \and
Yanfei Yao\inst{2}\textsuperscript{†} \and
Xuanbing Wen\inst{2} \and
Shixiong Zhang\inst{1}\and
Juli Zhang\inst{1}\textsuperscript{(\Letter)}\and
Kai Fan\inst{2}
}
\authorrunning{Z. Yan et al.}
%
\institute{
School of Computer Science and Technology, Xidian University, Xi’an 710126, Shaanxi, China\\
\email{22009200335@stu.xidian.edu.cn, zhangsx@xidian.edu.cn, zhangjuli@xidian.edu.cn}
\and
School of Cyber Engineering, Xidian University, Xi’an 710126, China\\
\email{d3we\_xidian@163.com, 22009200389@stu.xidian.edu.cn, kfan@mail.xidian.edu.cn}
}
\maketitle              
\let\thefootnote\relax\footnotetext{
\textsuperscript{†} Equal contribution.
}
%
\begin{abstract}

Vertical Federated Learning (VFL) has emerged as a promising paradigm for collaborative model training across distributed feature spaces, which enables privacy-preserving learning without sharing raw data. However, recent studies have confirmed the feasibility of label inference attacks by internal adversaries. By strategically exploiting gradient vectors and semantic embeddings, attackers-through passive, active, or direct attacks-can accurately reconstruct private labels, leading to catastrophic data leakage. Existing defenses, which typically address isolated leakage vectors or are designed for specific types of attacks, remain vulnerable to emerging hybrid attacks that exploit multiple pathways simultaneously. To bridge this gap, we propose Label-Anonymized Defense with Substitution Gradient (LADSG), a unified and lightweight defense framework for VFL. LADSG first anonymizes true labels via soft distillation to reduce semantic exposure, then generates semantically-aligned substitute gradients to disrupt gradient-based leakage, and finally filters anomalous updates through gradient norm detection. It is scalable and compatible with standard VFL pipelines. Extensive experiments on six real-world datasets show that LADSG reduces the success rates of all three types of label inference attacks by 30-60\% with minimal computational overhead, demonstrating its practical effectiveness.

\keywords{Vertical Federated Learning \and Label Inference Attack \and Gradient Substitution \and Label Anonymization \and Multi-Path Defense}
\end{abstract}
\section{Introduction}

Recent years have witnessed the remarkable potential of artificial intelligence (AI) in high-stakes domains such as healthcare and finance~\cite{miotto2018deep,heaton2017deep}. However, the growing dependency of AI models on high-quality and diverse datasets~\cite{Mohammed_2025,app13127082} has introduced significant challenges. Due to strict privacy regulations such as the General Data Protection Regulation (GDPR)~\cite{voigt2017gdpr} and the Health Insurance Portability and Accountability Act (HIPAA)~\cite{meingast2006hipaa}, critical data are often siloed across multiple organizations, making cross-institutional data integration difficult. This fragmentation, commonly referred to as the "data island" problem, severely hampers collaborative modeling and performance improvement~\cite{yang2019federated}.

Federated Learning (FL)~\cite{mcmahan2017communication} has emerged as a promising paradigm to address this structural contradiction. By enabling multiple parties to collaboratively train machine learning models without directly sharing raw data, FL offers a decentralized framework for learning from distributed data sources. It has demonstrated great potential in real-world applications such as mobile user behavior prediction and clinical diagnostics~\cite{hard2018federated,brisimi2018federated}.

According to the manner of data partitioning, FL can be broadly categorized into three types: \textit{Horizontal Federated Learning (HFL)}, \textit{Vertical Federated Learning (VFL)}, and \textit{Federated Transfer Learning (FTL)}~\cite{yang2019federated}. HFL is suitable for scenarios in which participating parties share the same feature space but possess different samples, typically across institutions with homogeneous data types. In contrast, VFL addresses cases where parties hold complementary feature spaces over a shared sample set, making it especially applicable to cross-industry or cross-organizational collaborations, such as jointly building user profiles between telecommunications providers and financial institutions. FTL is designed for situations where both the sample space and the feature space differ significantly among parties, leveraging transfer learning techniques to enable cross-domain collaborative modeling.

Although raw data are not directly shared in FL, the process of feature computation and exchange-combined with the risk of potential participant compromise-can still lead to privacy leakage at the feature level~\cite{vucinich2023current,lyu2022privacyrobustnessfederatedlearning}. Against this backdrop, \textit{label inference attacks} pose a critical threat as they can expose highly sensitive user attributes (e.g., medical conditions or financial status). This paper builds on the attack framework proposed by Fu et al.~\cite{fu2022label} in the context of VFL, which demonstrates that adversaries can accurately reconstruct target labels by leveraging (i) their locally trained models and (ii) server-returned gradient updates. The attack strategies can be broadly classified into three types: \emph{passive attacks}, \emph{active attacks}, and \emph{direct attacks} (see §\ref{sec:attack-categorization} for details). Existing defenses can be broadly categorized based on the targeted leakage pathway: (1) \textit{model-side defenses} such as KDK~\cite{arazzi2024kdk}, LabObf~\cite{he2024labobflabelprotectionscheme}, and VAFL~\cite{chen2020vafl} mitigate label exposure via perturbation or pseudolabeling; (2) \textit{gradient-side} defenses like SignSGD~\cite{bernstein2018signsgd}, CAFE~\cite{jin2021cafe}, DP-FL~\cite{kairouz2019advances}, and PPDL~\cite{shokri2015privacy} intervene through clipping, noise injection, or gradient obfuscation. However, most existing approaches suffer from two key limitations: they (i) focus solely on either model-side or gradient-side leakage, and thus fail to address scenarios where both leakage pathways are exploited jointly~\cite{qiu2022your}; and (ii) are tailored to specific attack types (e.g., passive or active attacks), rather than providing a unified and generalizable defense against label inference, which ultimately limits their robustness and adaptability.

To bridge this gap, we propose a novel defense framework, \textbf{LADSG} (\textit{Label-Anonymized Defense with Substitution Gradient}), which provides unified protection against both \textit{model-side} and \textit{gradient-side} label leakage in VFL. Unlike most existing methods that target a single leakage path, LADSG is the first framework to jointly address both attack surfaces within an integrated architecture. It designs a multi-layered defense architecture composed of three complementary server-side modules. First, to mitigate model-side semantic leakage-where adversaries exploit the local encoder’s learned semantics to infer labels-we introduce the \textbf{Label-Anonymized Distillation (LADistill)} module. Instead of exposing hard labels, LADistill generates soft labels using a lightweight server-side teacher model, which preserves informative class distributions while reducing label specificity. To stabilize training and mitigate noise from low-confidence predictions, a top-k filtering strategy is applied to retain only the most probable classes in each soft label output. Second, to prevent gradient-side label leakage-where attackers exploit server-returned gradients to infer private labels, we propose the \textbf{Similar Gradient Substitution (SGSub)} module. It generates fake gradients sampled from a Gaussian distribution and selects substitutes via a dual similarity constraint based on Mahalanobis distance and cosine similarity. This ensures statistical indistinguishability while avoiding label-sensitive directions, balancing gradient plausibility and privacy. Finally, to defend against active adversaries that manipulate the training process to bypass defense mechanisms, we employ the \textbf{Gradient-based Anomaly Detection (GENO)} module. It continuously monitors gradient norms and distributional shifts during training, enabling effective detection of adversarial behaviors (e.g., learning rate manipulation), thereby enhancing robustness and privacy protection. Experiments demonstrate that LADSG reduces the accuracy of all three types of label inference attacks by 30\%–60\% compared to the no-defense baseline.

In summary, our main contributions are as follows:
\begin{itemize}
    \item We propose a unified defense strategy targeting multiple label inference attacks introduced by Fu \textit{et al.}~\cite{fu2022label}. To the best of our knowledge, this is the first framework that jointly defends against both local model leakage and gradient-based leakage.

    \item We conduct a comprehensive evaluation of LADSG across six real-world datasets spanning vision, language, healthcare, and finance domains. Results demonstrate that our method consistently reduces attack accuracy across all attack types, and outperforms existing defenses in both robustness and effectiveness.

    \item We incorporate the t-SNE visualization technique~\cite{van2008visualizing} to analyze the distribution of model output features, providing further insight into how LADSG mitigates label leakage.
\end{itemize}
\section{Related Work}

Recent studies have shown that FL is vulnerable to various inference attacks, including \textit{membership inference}, \textit{property inference}, and \textit{feature inference}~\cite{melis2019exploiting,nasr2018comprehensive,zhu2019deep}. 
Membership inference aims to determine whether a particular sample was part of a participant's training set. However, this threat is less relevant in \textit{VFL}, where sample identifiers are shared across parties by design. Property inference attacks aim to recover feature attributes that are unrelated to the main learning task, while feature inference attacks attempt to reconstruct private features held by other participants. However, these attacks primarily target the reconstruction of input features, rather than exposing the semantic information associated with the model’s outputs, such as label predictions.

Unlike the aforementioned inference attacks, \textit{label inference attacks} have attracted increasing attention in recent years due to the highly sensitive nature of label information (e.g., disease status, income level). Several studies~\cite{li2021label,liu2024similarity,liu2021batch,sun2022label} have proposed various attack strategies under different collaborative learning settings, such as clustering on smashed data or gradient inversion, demonstrating that accurate label recovery is possible even without explicit label sharing. In the VFL setting, Fu \textit{et al.}~\cite{fu2022label} identify three types of label inference attacks~(see §\ref{sec:attack-categorization}), revealing two fundamental leakage channels: (i) \textbf{Model path} - semantic cues embedded in the attacker's local model; (ii) \textbf{Gradient path} - implicit label signals encoded in the backpropagated gradients returned by the server. 
These two channels form the core vulnerabilities exploited by label inference attacks in VFL, and directly motivate the design of our defense framework.

Existing label privacy defense methods can be broadly categorized based on the attack origin into the following two classes: \textit{(1) Model-path defenses.} These methods aim to prevent attackers from inferring label semantics via their local models, often by applying label obfuscation or representation perturbation. For example, KDk~\cite{arazzi2024kdk} combines soft-label distillation with \(k\)-anonymity to anonymize labels, but requires a large teacher model for effectiveness. LabObf~\cite{he2024labobflabelprotectionscheme} enhances uncertainty through soft-label mapping, yet incurs significant performance overhead when dealing with large label spaces. VAF-L~\cite{chen2020vafl} employs asynchronous updates to mitigate label leakage, but suffers from gradient staleness and residual leakage paths that compromise convergence and privacy. \textit{(2) Gradient-path defenses.} This line of work focuses on preventing attackers from inferring labels through gradients returned by the server, and generally follows one of three strategies: sensitive gradient filtering, perturbation mechanisms, or synthetic gradient generation. For instance, PPDL~\cite{shokri2015privacy} filters sensitive gradients centrally, but relies on prior identification of sensitive components and lacks adaptability to evolving threats.  
Abadi \textit{et al.}~\cite{abadi2016deep} apply differential privacy via noise injection, introducing a privacy–utility trade-off.  
Top-$k$ sparsification~\cite{kairouz2019advances} and sign quantization~\cite{bernstein2018signsgd} compress gradients to reduce sensitivity, yet risk discarding critical structures, compromising convergence. More recently, CAFE~\cite{jin2021cafe} and FLSG~\cite{fan2024flsg} enhance obfuscation by generating substitute gradients based on cosine similarity matching. However, this naive replacement strategy often fails to disrupt class-discriminative patterns, making the synthetic gradients easily distinguishable from genuine ones~\cite{Li_2021,hitaj2017deep}.

It is important to note that most existing defense methods assume the attacker exploits only a single leakage channel, overlooking a more realistic and severe threat, where the adversary simultaneously leverages both the semantic information in their local model and the gradient signals returned by the server for label inference. Recent studies~\cite{wang2024breakingsecureaggregationlabel,zhang2024buildinggradientbridgeslabel,Gat2024harmful} have demonstrated that such joint inference attacks can significantly improve attack accuracy and effectively bypass single-path defenses. As a result, current defenses exhibit blind spots when faced with these compound threats. To address these limitations, we propose \textbf{LADSG}, a unified defense framework that systematically addresses label leakage through both the model and gradient channels. This design substantially enhances the robustness of VFL systems against label inference attacks.
\section{Preliminaries}

\subsection{Problem Setting}

\begin{table}[t]
\centering
\caption{Notation summary for vertical federated learning}
\begin{tabular}{ll}
\toprule
\textbf{Symbol} & \textbf{Description} \\
\midrule
\(K\) & Number of participating entities in VFL \\
\(E_k\) & The \(k\)-th entity (data owner) \\
\(n\) & Number of aligned users across all entities \\
\(u_i\) & The \(i\)-th user (\(i = 1, \dots, n\)) \\
\(x_k^{(i)}\) & Feature vector of user \(u_i\) held by entity \(E_k\) \\
\(X_k \in \mathbb{R}^{n \times d_k}\) & Local dataset at entity \(E_k\) with feature dimension \(d_k\) \\
\(y^{(i)}\) & Label of user \(u_i\), owned by the label owner \\
\(Y \in \mathbb{R}^{n \times c}\) & Label matrix (\(c\) is number of classes or 1 for regression) \\
\(f_k(\cdot; \theta_k)\) & Local model at entity \(E_k\), parameterized by \(\theta_k\) \\
\(h(\cdot; \theta_h)\) & Aggregation and prediction head at label owner \\
\(\hat{y}^{(i)}\) & Predicted label for user \(u_i\) \\
\(\ell(\cdot, \cdot)\) & Task-specific loss function (e.g., cross-entropy) \\
\(\mathcal{L}\) & Overall training loss over all users \\
\bottomrule
\end{tabular}
\label{tab:notation}
\end{table}

In this paper, we consider a VFL framework involving $K$ entities denoted as $\{ E_1, E_2, ..., E_K \}$. Each entity $E_K$ holds a local feature subset $X_k \in \mathbb{R}^{n \times d_k}$ corresponding to a common set of users, with user identities aligned across entities. Among them, a designated label owner possesses the ground-truth labels $Y$. The collaborative objective is to jointly learn a predictive model that maps the concatenated features $(X_1, X_2, ..., X_K)$ to the labels $Y$, without exposing raw data across entities. 

Let  $f_k(\cdot; \theta_k)$ denote the local encoder of entity $E_K$, and $h(\cdot; \theta_k)$ indicate the prediction head at the label owner. For a given sample, the final prediction is defined as follows:
\begin{equation}
\hat{y} = h\left(f_1(x_1), f_2(x_2), \dots, f_K(x_K)\right).
\end{equation}
The training objective is to jointly optimize the local model parameters \(\{\theta_k\}\) and the prediction head parameters \(\theta_h\), by minimizing the supervised loss:
\begin{equation}
 \mathcal{L}(\{\theta_k\}, \theta_h) = \mathbb{E}_{(x_1, \dots, x_K, y) \sim \mathcal{D}} \left[ \ell\left( h(f_1(x_1; \theta_1), \dots, f_K(x_K; \theta_K); \theta_h), y \right) \right],   
\end{equation}
where \(\ell(\cdot, \cdot)\) denotes the task-specific loss function. Each local model \(f_k(\cdot; \theta_k)\) is parameterized by \(\theta_k\).

During training, each entity computes local representations based on its features. These representations are sent to the label owner, which aggregates them to predict labels and compute the loss. Gradients are then exchanged among entities to update local models. To facilitate clarity, we summarize the key notations used throughout this paper in Table~\ref{tab:notation}.

\subsection{Threat Model: Label Inference Attacks}
\label{sec:attack-categorization}
\begin{figure}[t]
    \centering
    \includegraphics[width=0.8\linewidth]{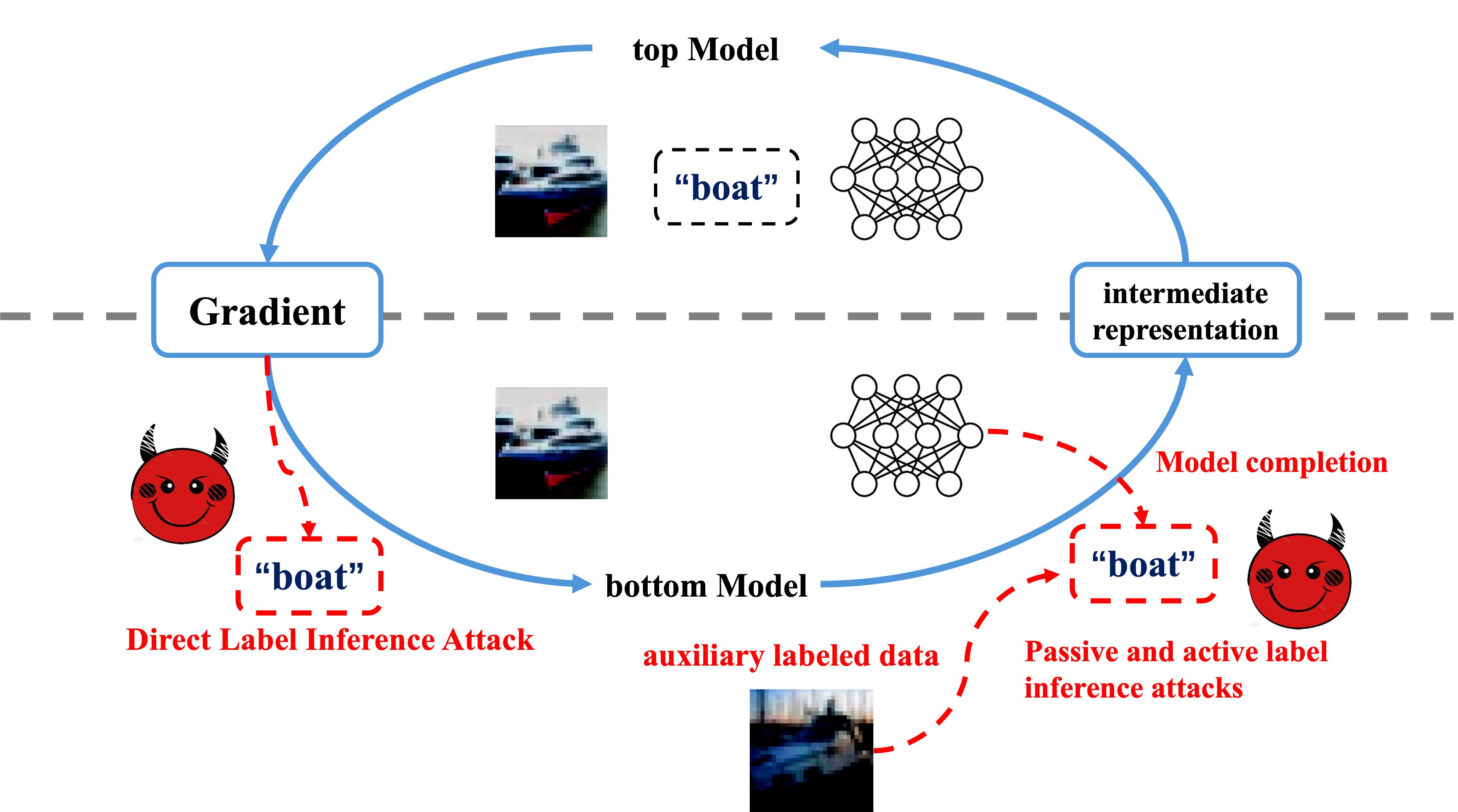}
    \caption{Taxonomy of label inference attacks in VFL. Passive attacks analyze embeddings; active attacks distort optimization; 
    direct attacks invert gradients.}
    \label{fig:label inference attacks}
\end{figure}

Following the categorization proposed by Fu \textit{et al.}~\cite{fu2022label}, we divide label inference attacks in VFL into three types: \textit{Passive}, \textit{Active}, and \textit{Direct} attacks.  
These attack strategies differ in how they exploit the two known leakage channels-model outputs and gradient updates-and are summarized in Figure~\ref{fig:label inference attacks}.  
We provide a detailed analysis of each type in the following subsections.

\paragraph{Passive Attacks.} 
In passive attacks, the adversary leverages its locally held model to infer labels. These attacks are termed "passive" because the attacker adheres to the protocol during training and behaves as an honest-but-curious participant. By observing intermediate representations produced by the local model, the attacker extracts semantic information. It is typically assumed that the attacker has access to a small amount of auxiliary labeled data (e.g., 0.08\% of the training labels as used in~\cite{fu2022label}) and attaches a classifier or fine-tunes its model in a semi-supervised fashion to recover partial label information.

\paragraph{Active Attacks.} 
Active attacks involve intentional manipulation of the local optimization process during training to amplify the representation of label-relevant features. The attacker may enlarge local gradients (e.g., by adjusting the learning rate or employing malicious optimizers), forcing the top model to become more reliant on its input. After training, the attacker fine-tunes its model using a small amount of auxiliary labels to recover the true labels more accurately.

\paragraph{Direct Attacks.} 
Direct attacks exploit the gradients returned by the top model to directly infer the ground-truth labels of training samples. These attacks typically analyze the sign of the loss gradients and do not require access to model internals. Notably, such attacks remain effective in standard VFL settings without explicit model partitioning, and have been validated both theoretically and empirically.

These attack strategies highlight that both local model outputs and backpropagated gradients constitute major leakage vectors for label inference, necessitating a unified defense against both paths.
\section{Methodology}

\begin{figure}[t]
    \centering
    \includegraphics[width=0.75\linewidth]{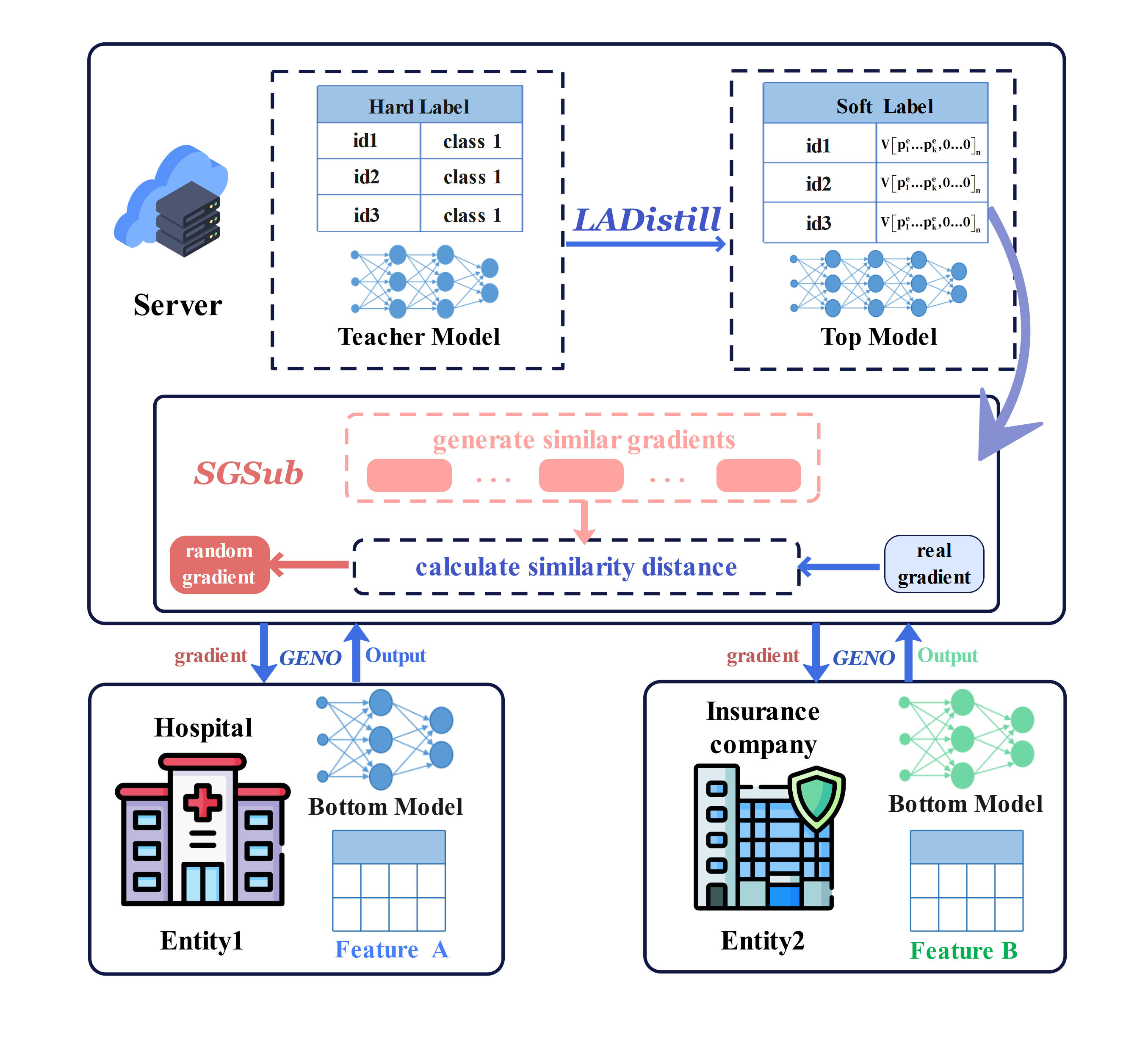}
    \caption{Overview of LADSG. Our framework combines gradient substitution (SGSub), label anonymization (LADistill), and anomaly detection (Geno) to jointly mitigate diverse label inference threats in VFL.}
    \label{fig:our_work}
\end{figure}

In this paper, the proposed \textbf{LADSG} is a unified framework composed of three collaborative modules: \textit{Similar Gradient Substitution (SGSub)}, \textit{Label-Anonymized Distillation (LADistill)}, and \textit{Gradient-based Anomaly Detection (GENO)}. They form an end-to-end defense pipeline that preserves model utility while countering the three types of label inference threats defined by Fu \textit{et al.}~\cite{fu2022label}. Figure~\ref{fig:our_work} shows the overall architecture of LADSG. Next, we detail the design and implementation of each module.

\subsubsection{Similar Gradient Substitution}

To defend against gradient-based label inference, we introduce \textbf{SGSub}, a constrained gradient substitution mechanism. Instead of directly returning sensitive gradients to each participant, SGSub replaces them with synthetic surrogates that preserve both directional and distributional similarity.

SGSub improves upon prior surrogate methods such as FLSG~\cite{fan2024flsg} by enforcing a dual similarity constraint: sampled gradients must align in cosine direction and match the statistical shape (via Mahalanobis distance) of the original vector. This ensures that substituted gradients are indistinguishable from real ones, even under distribution-aware inference~\cite{Li_2021,hitaj2017deep}.

Specifically, at each training iteration, the server computes the original gradient vector \( g_k \) and samples candidate vectors \( \hat{v} \sim \mathcal{N}(\mu, \varphi^2) \), where the mean \( \mu \) and variance \( \varphi^2 \) are estimated from historical gradient statistics across training rounds. We adopt Gaussian sampling as it empirically approximates the distribution of gradient updates in deep learning models~\cite{fan2024flsg}. This approach provides an efficient and controllable way to generate plausible surrogates that fall within the statistical envelope of previously observed gradients.

To assess the fidelity of each candidate \( \hat{v} \), we define a combined similarity score against the true gradient \( v \) as:
\begin{equation}
\text{Sim}(\hat{v}, v) = W_{\text{Cos}} \cdot \text{CosSim}(\hat{v}, v) + W_M \cdot D_M(\hat{v}, v) \leq \tau
\end{equation}
Here, the term \( \text{CosSim}(\hat{v}, v) = \frac{\hat{v} \cdot v}{\|\hat{v}\|_2 \|v\|_2} \) denotes the cosine similarity, which meanseures the angular alignment between \( \hat{v} \) and \( v \). In contrast, \( D_M(\hat{v}, v) = \sqrt{(\hat{v} - v)^T S^{-1} (\hat{v} - v)} \) is the Mahalanobis distance, which quatifies the dissimilarity in terms of statistical distribution under the covariance matrix \( S \). \( W_{\text{Cos}} \) and \( W_M \) are the scalar weights that balance the contribution of each similarity, and \( \tau \) is the acceptance threshold.

Only candidates satisfying the similarity condition (\( \text{Sim}(\hat{v}, v) \leq \tau \)) are selected. Once accepted, the surrogate gradient vector is reordered to match the original index structure of \( v \) and returned to the corresponding participant.

The full procedure is outlined in Algorithm~\ref{alg:sgsub}, which integrates stochastic sampling, value clamping, similarity filtering, and structural reordering.

\begin{algorithm}[ht]
\caption{SGSub: Similar Gradient Substitution}
\label{alg:sgsub}
\small
\begin{algorithmic}[1]
\Require Top model $\theta_{\mathrm{top}}$, threshold $\tau$, similarity weights $W_{\text{Cos}}, W_M$
\Ensure Privacy-preserving surrogate gradients $\hat{g}_k^i$ for each participant $k$

\For{each iteration $i = 1,2,\dots$}
    \State Sample a mini-batch of size $b$
    \State Receive $o_1^i, \dots, o_K^i$ from all participants
    \State $o^i_{\mathrm{all}} \gets \mathrm{Concat}(o_1^i, \dots, o_K^i)$
    \State Compute loss $\mathcal{L} \gets \mathrm{CrossEntropy}(\theta_{\mathrm{top}}(o^i_{\mathrm{all}}), Y)$
    \For{each participant $k=1,2,\dots,K$}
        \State $g_k^i \gets \frac{\partial \mathcal{L}}{\partial o_k^i}$
        \State Flatten $g_k^i$ to $v$, record sort index $\xi$
        \Repeat
            \State Sample $\hat{v} \sim \mathcal{N}(\mu, \varphi^2)$ and clamp to $(\min(v), \max(v))$
            \State Compute similarity $s = W_{\text{Cos}} \cdot \text{CosSim}(\hat{v}, v) + W_M \cdot D_M(\hat{v}, v)$
        \Until{$s \leq \tau$ or maximum attempts reached}
        \State $\hat{g}_k^i \gets \mathrm{Reorder}(\hat{v}, \xi)$
        \State Return $\hat{g}_k^i$ to participant $k$
    \EndFor
\EndFor
\end{algorithmic}
\normalsize
\end{algorithm}

\subsubsection{Label-Anonymized Distillation}

To mitigate model-path label leakage, we propose \textbf{LADistill}, a \emph{deterministic} label anonymization mechanism that combines soft-label distillation with $k$-anonymity filtering. LADistill is designed to obscure class-discriminative signals that may be exploited by adversaries while preserving the convergence behavior of the model during training.

The overall LADistill pipeline consists of three components: (1) soft label generation via a self-distilled teacher model, (2) deterministic $k$-anonymity filtering to suppress overconfident class indicators, and (3) a modified training objective based on anonymized soft targets. We describe each step in detail below.

\paragraph{Soft-Label Generation}  
We first construct a lightweight teacher network to generate softened probabilistic labels. This teacher model is initialized using an exponential moving average (EMA) of the main model’s weights and updated continuously via self-distillation on the same task. Given a hard labels \(\mathrm{HL}\), the teachter projects it into a class probability distributions \(\mathrm{SL}\).  
For the \(i\)-th sample with logits \(\mathbf{z}_i \in \mathbb{R}^C\), the soft label is computed as:
\begin{equation}
    \mathrm{SL}_{ij}= \frac{\exp(z_{ij})}{\sum_{c=1}^{C}\exp(z_{ic})},\qquad j=1,\dots,C,
\end{equation}
where \(C\) denotes the number of classes. This softening (with temperature=1) distributes the gradient source across all the classes, reather than focusing it on a single one, thereby reducing gradient variance and enhancing convergence stability, as widely demonstrated in prior work~\cite{gou2021knowledge}.

\paragraph{$k$-Anonymity Filtering}  
To further suppress the identifiability of the highest-probability class, we apply controlled obfuscation to \(\mathrm{SL}\).  
Let \(\mathcal{T}_k(i) = \mathrm{Top}\text{-}k(\mathrm{SL}_i)\) denote the indices of the top-\(k\) predicted classes for sample \(i\), the anonymized label \(\hat{\mathrm{SL}}_{ij}\) is defined as:
\begin{equation}
    \hat{\mathrm{SL}}_{ij}=
\begin{cases}
1-\epsilon, & j=\arg\max \mathrm{SL}_i,\\[2pt]
\dfrac{\epsilon}{k-1}, & j \in \mathcal{T}_k(i) \!\setminus\! \{\arg\max \mathrm{SL}_i\},\\[6pt]
0, & \text{otherwise}.
\end{cases}
\end{equation}
Here, \( \epsilon \in (0, 1) \) controls the privacy--utility trade-off. Since \( \hat{\mathrm{SL}} \) shares the same support set \( \mathcal{T}_k(i) \) as \( \mathrm{SL} \), it introduces no stochastic noise; instead, it deterministically redistributes the label probability mass within the top-\(k\) candidates. This approach effectively extends the concept of label smoothing while constraining it to \( \mathcal{T}_k \), thereby significantly reducing information loss.

\paragraph{Training Objective}  
The student network is trained to minimize the cross-entropy between its predictions and the anonymized soft labels:
\begin{equation}
    \mathcal{L}_{\text{LAD}} = \mathrm{CE}\!\bigl(\mathrm{Softmax}(\mathbf{p}),\,\hat{\mathrm{SL}}\bigr)
\end{equation}

\begin{algorithm}[ht]
\caption{LADistill: Label-Anonymized Distillation}
\label{alg:lad}
\small
\begin{algorithmic}[1]
\Require Input $x$, teacher model $f_t$, smoothing parameter $\epsilon$, top-$k$ value $k$
\Ensure Anonymized soft label vector $\hat{SL}$

\State $z \gets f_t(x)$
\State $SL \gets \mathrm{softmax}(z)$
\State $(TopKIndexes, MaxIndex) \gets \mathrm{getTopK}(SL, k)$
\State $\hat{SL} \gets \mathrm{zeros}(n)$
\ForAll{$i$ in $TopKIndexes$}
    \If{$i == MaxIndex$}
        \State $\hat{SL}[i] \gets 1 - \epsilon$
    \Else
        \State $\hat{SL}[i] \gets \epsilon / (k - 1)$
    \EndIf
\EndFor
\State \Return $\hat{SL}$
\end{algorithmic}
\normalsize
\end{algorithm}

where \(\mathbf{p}\) denotes the student logits.  
As gradients are derived from probability distributions rather than hard labels, \(\mathcal{L}_{\text{LAD}}\) remains smooth and continuous w.r.t. \(\epsilon\) and \(k\), thus exhibiting comparable convergence behavior to standard cross-entropy without introducing instability. The full procedure is illustrated in Algorithm~\ref{alg:lad}.

\subsubsection{Gradient-Based Anomaly Detection}

SGSub and LADistill assume that all participants are honest-but-curious. To guard against stronger adversaries-such as those intentionally manipulating their local training dynamics (e.g., amplifying learning rates)-we introduce a lightweight runtime module, \textbf{GENO}, which detects anomalies solely by monitoring the \( L_2 \) norm of uploaded gradients.

\begin{algorithm}[ht]
\caption{\scriptsize Geno: Gradient-Based Evaluation of Network Outliers}
\label{alg:geno}
\small
\begin{algorithmic}[1]
\Require Gradient set $G$, threshold $\lambda$
\Ensure Filtered gradient set $G_{\mathrm{filtered}}$
\State $G_{\mathrm{filtered}} \gets \emptyset$
\ForAll{$g_i \in G$}
    \If{$\|g_i\|_2 \leq \lambda$}
        \State $G_{\mathrm{filtered}} \gets G_{\mathrm{filtered}} \cup \{g_i\}$
    \EndIf
\EndFor
\State \Return $G_{\mathrm{filtered}}$
\end{algorithmic}
\end{algorithm}

Unlike Zeno~\cite{Xie2019Zeno}, which requires additional communication to compare full model updates, GENO operates without extra overhead: in each round, it simply discards gradients that exceed a predefined norm threshold: $\lVert\mathbf{g}_i\rVert_2 > \lambda$ and retains the remaining ones:
\begin{equation}
\mathcal{G}_{\text{filtered}} = \left\{ \mathbf{g}_i \in \mathcal{G} \,\middle|\, \lVert\mathbf{g}_i\rVert_2 \le \lambda \right\},
\end{equation}

where \( \mathbf{g}_i \) denotes the gradient from a single party and \( \mathcal{G} \) is the set of all gradients in the current round.  
The threshold \( \lambda \) is initialized once at the beginning of training based on a clean baseline and remains fixed throughout.  
Since only a small fraction of extreme values are removed, GENO introduces negligible impact on model convergence. The full procedure is summarized in Algorithm~\ref{alg:geno}.
\section{Experiment and Discussion}

Our evaluation of LADSG aims to address the following key questions:

\begin{itemize}
\item How effective is LADSG at defending against the three major types of label inference attacks, compared to state-of-the-art defenses? (§\ref{Defense Effectiveness})
\item How does LADSG affect the performance of the underlying model relative to existing defense methods? (§\ref{Model Performance and Runtime Efficiency})
\item What are the individual and combined contributions of LADSG’s three components to its overall defense effectiveness? (§\ref{Ablation Study})
\end{itemize}

\subsection{Experimental Setup}

\subsubsection{Datasets and Model Architectures}

We evaluate LADSG on six real-world datasets from diverse domains. Table~\ref{tab:datasets} summarizes the dataset statistics and model configurations.

\begin{table}[ht]
\centering
\caption{Datasets and Model Architecture}
\label{tab:datasets}
\begin{tabular}{lcccc c}
\toprule
\textbf{Dataset} & \textbf{Train Size} & \textbf{Test Size} & \textbf{Class} & \textbf{Bottom Model} & \textbf{Top Model} \\
\midrule
CIFAR-10         & 50,000   & 10,000  & 10   & ResNet-20 & FCNN-4 \\
CIFAR-100        & 50,000   & 10,000  & 100  & ResNet-20 & FCNN-4 \\
CINIC-10         & 180,000  & 90,000  & 10   & ResNet-20 & FCNN-4 \\
Yahoo Answers    & 50,000   & 20,000  & 10   & BERT      & FCNN-3 \\
Loan Prediction  & 80,000   & 20,000  & 2    & FCNN-4    & FCNN-3 \\
BHI              & 69,000   & 17,000  & 2    & ResNet-20 & FCNN-4 \\
\bottomrule
\end{tabular}
\end{table}

CIFAR-10~\cite{krizhevsky2009learning}, CIFAR-100~\cite{krizhevsky2009learning}, and CINIC-10~\cite{darlow2018cinic} are standard image classification benchmarks with $32 \times 32$ color images. CIFAR-100 includes 100 categories, while CINIC-10 provides a large-scale testbed with 270{,}000 samples drawn from CIFAR and ImageNet. These datasets are used to evaluate LADSG's scalability and robustness in visual tasks. Yahoo Answers~\cite{zhang2015character} is a large-scale NLP dataset consisting of question-answer pairs, used to assess LADSG in sentence-level text classification where token sequences are vertically partitioned across participants. The Loan Default Prediction dataset~\cite{loanprediction} includes 156 tabular features for financial record analysis, simulating enterprise VFL settings. BHI~\cite{janowczyk2016deep} is a breast histopathology dataset for binary cancer diagnosis, representing high-stakes medical applications.

For all image tasks, we adopt ResNet-20~\cite{he2016deep} as the bottom model for each participant and use a four-layer fully connected top model (FCNN-4) at the server. The input images are vertically split into two halves and assigned to two participants. For text and tabular tasks, we use BERT~\cite{devlin2019bert} or lightweight fully connected networks as bottom models, with a smaller top model (e.g., FCNN-3) deployed on the server. In the text setting, each participant holds a different segment of the same sentence or paragraph.

\begin{table}[ht]
    \centering
    \caption{Comparison of Teacher Model Architectures in LADistill vs. KDk}
    \label{tab:teacher_models}
    \begin{tabular}{lcc}
        \toprule
        \textbf{Dataset} & \textbf{LADistill (Ours)} & \textbf{KDk~\cite{arazzi2024kdk}} \\
        \midrule
        CIFAR-10 & ResNet-20 + FCNN-1 & ResNet-50 + FCNN-1 \\
        CIFAR-100 & ResNet-20 + FCNN-1 & ResNet-50 + FCNN-1 \\
        CINIC-10 & ResNet-20 + FCNN-1 & ResNet-50 + FCNN-1 \\
        Yahoo Answers & Distilled BERT + FCNN-1 & BERT + FCNN-1 \\
        Loan Prediction & FCNN-3 & \textbackslash \\
        BHI & FCNN-3 & \textbackslash \\
        \bottomrule
    \end{tabular}
\end{table}

Table~\ref{tab:teacher_models} compares the teacher network architectures used in LADistill and KDk~\cite{arazzi2024kdk}. Both methods utilize soft-label distillation with $k$-anonymity, but LADistill employs significantly lighter teacher models, such as shallower CNNs and distilled versions of BERT. This design choice reduces computational overhead and enhances deployability, particularly in resource-constrained VFL environments.

\subsubsection{Evaluation Metrics}
Following prior work~\cite{fan2024flsg,arazzi2024kdk}, we adopt attack success rate and model accuracy as the primary evaluation metrics to assess the model’s ability to resist label inference while maintaining utility. The specific evaluation criteria vary based on task type and label distribution. For datasets with a small number of classes (e.g., CIFAR-10, CINIC-10, and Yahoo Answers), we report Top-1 accuracy. For CIFAR-100, which has a larger class set, we use Top-5 accuracy to better capture prediction capability. For highly imbalanced datasets such as BHI and Loan Default Prediction, we report the F1 score to more effectively reflect class-wise robustness and attack resistance.

\subsubsection{Baseline Methods}
We compare LADSG against six representative defense methods that cover the main strategies currently used to mitigate label inference attacks: (1) \textbf{Gradient Clipping (GC)}~\cite{kairouz2019advances}: Retains a fixed proportion of high-magnitude gradients to reduce sensitive information leakage and communication cost; (2) \textbf{Gradient Noise (NGs)}~\cite{zhu2019deep}: Adds Laplacian noise to backpropagated gradients to perturb label-related signals; (3) \textbf{Multi-bucket Gradient Quantization (MGs)}~\cite{bernstein2018signsgd}: An extension of SignSGD that partitions gradients based on variance to achieve efficient compression and privacy enhancement; (4) \textbf{Privacy-Preserving Deep Learning (PPDL)}~\cite{shokri2015privacy}: Combines selective gradient sharing and random update strategies to improve resistance to inference; (5) \textbf{Fake Gradient Substitution (FLSG)}~\cite{fan2024flsg}: Synthesizes surrogate gradients to replace real ones, thereby disrupting label recovery; (6) \textbf{Label Distillation (Kdk)}~\cite{arazzi2024kdk}: Employs a teacher–student paradigm to eliminate direct exposure of labels during training and weaken label leakage at its source.

These methods reflect a range of core defense strategies, including noise injection, gradient quantization, substitution-based masking, and training path restructuring.  
To ensure a fair comparison, all baselines are evaluated under identical data splits, model architectures, and training procedures.

\subsubsection{Experimental Environment}

All experiments were conducted on a workstation with an AMD Ryzen Threadripper 3960X CPU, 64GB RAM, and an NVIDIA GeForce RTX 3090 GPU. We implement all methods using PyTorch and utilize GPU acceleration for model training and inference.

\subsection{Defense Effectiveness}
\label{Defense Effectiveness}

\begin{figure}[t]
    \centering
    \includegraphics[width=\textwidth]{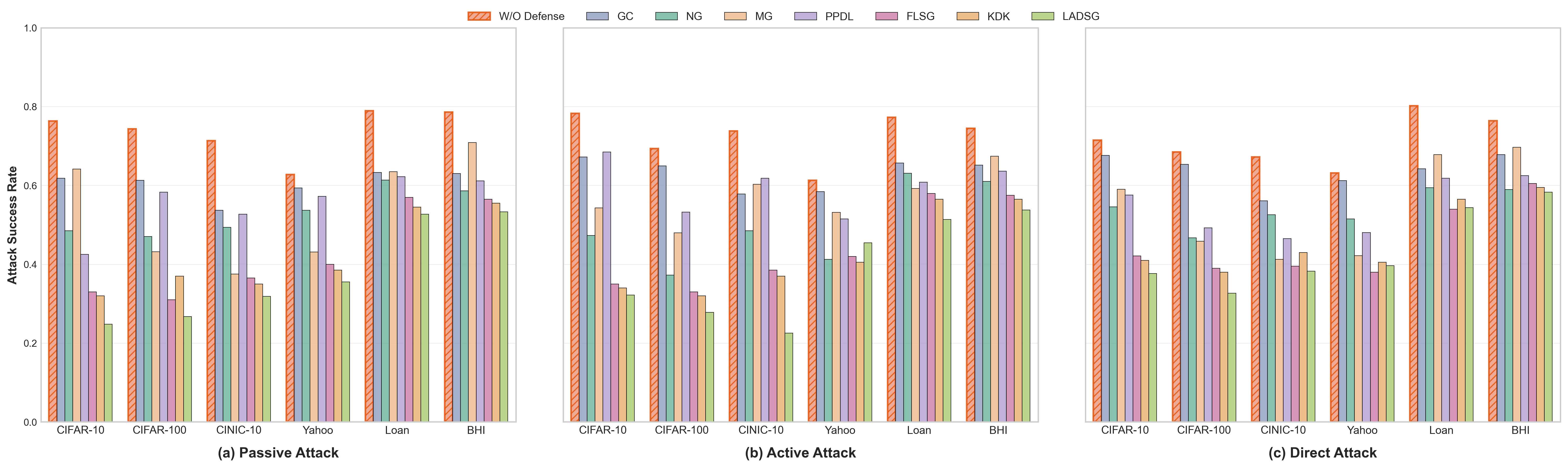}
    \caption{Defense effectiveness of LADSG under three types of label inference attacks across six datasets.}
    \label{fig:attack_comparison}
\end{figure}

In this experiment, we systematically evaluate eight defense methods across six datasets under three types of label inference attacks: passive, active, and direct. For each defense, we report the attack success rate using its optimal parameter configuration. Figure~\ref{fig:attack_comparison} compares the performance of all methods under different attack scenarios. We highlight three key findings: (i) traditional defenses fail to generalize across attack types, (ii) recent targeted methods remain vulnerable under complex threats, and (iii) our proposed LADSG achieves the most robust and consistent performance by integrating multi-path defenses.

First, all three attack types achieve high success rates when no defense is applied($0.70 \sim 0.80$), revealing VFL’s vulnerability to label inference. While traditional defenses (GC, NG, MG, PPDL) partially reduce attack success rates, they lack adaptability. For instance, PPDL is ineffective against active attacks, and GC fails to defend against direct attacks. These limitations are due to their reliance on single-strategy defenses-e.g., GC only truncates gradient magnitudes but retains directional cues; NG adds unstructured noise, easily smoothed during averaging; MG and PPDL reduce communication bandwidth but not label-relevant directions. Second, more recent defenses like FLSG and KDK introduce targeted strategies-FLSG perturbs gradients to mimic original directions, mitigating gradient-path leakage; KDK combines k-anonymity with label distillation to reduce the semantic leakage of true labels from local models. However, both still show non-negligible attack success rates (30--40\%), indicating insufficient protection under multi-path adversaries. 

In contrast, \textbf{LADSG} consistently achieves the lowest attack success rates across all settings, outperforming the best baseline by 10--20\%. This is attributed to its tri-level architecture: \emph{LADistill} compresses label semantics to minimize embedding-label mutual information; \emph{SGSub} perturbs both gradient magnitude and sign via dual statistical constraints; \emph{GENO} detects gradient anomalies in real-time to filter active manipulations. These modules jointly secure model-path, gradient-path, and behavior-path, forming a comprehensive defense that hinders inference through any single vector.

\begin{figure}[ht]
    \centering
    \resizebox{0.75\textwidth}{!}{%
        \begin{minipage}[c]{0.29\textwidth}
            \centering
            \includegraphics[width=0.95\linewidth]{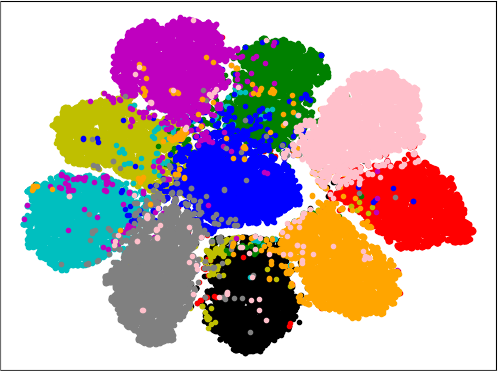}
            \caption*{(a)}
        \end{minipage}
        \hspace{0.02\textwidth}
    \begin{minipage}[c]{0.02\textwidth}
        \centering
        \parbox{0.4cm}{
            \rule{0.4pt}{0.2cm}\\[0.10cm]
            \rule{0.4pt}{0.2cm}\\[0.10cm]
            \rule{0.4pt}{0.2cm}\\[0.10cm]
            \rule{0.4pt}{0.2cm}\\[0.10cm]
            \rule{0.4pt}{0.2cm}\\[0.10cm]
            \rule{0.4pt}{0.2cm}\\[0.10cm]
            \rule{0.4pt}{0.2cm}\\[0.10cm]
        }
    \end{minipage}
        \hspace{0.02\textwidth}
        \begin{minipage}[c]{0.55\textwidth}
            \centering
            \begin{minipage}[c]{0.32\textwidth}
                \centering\includegraphics[width=\linewidth]{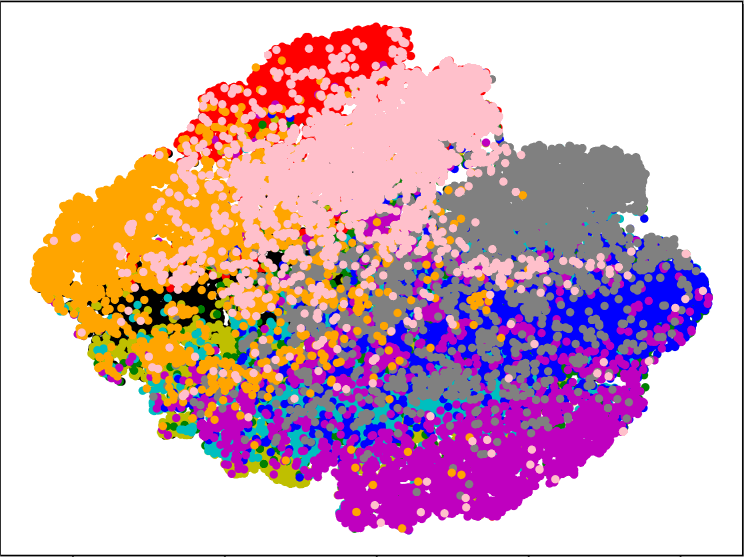}
                \caption*{(b)}
            \end{minipage}\hfill
            \begin{minipage}[c]{0.32\textwidth}
                \centering\includegraphics[width=\linewidth]{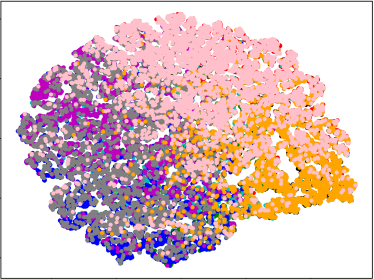}
                \caption*{(c)}
            \end{minipage}\hfill
            \begin{minipage}[c]{0.32\textwidth}
                \centering\includegraphics[width=\linewidth]{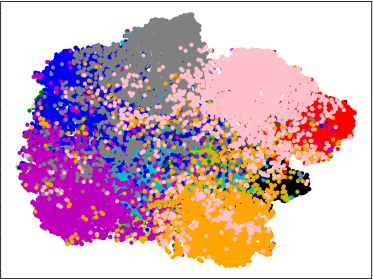}
                \caption*{(d)}
            \end{minipage}
            \vspace{0.12cm} 
            \begin{minipage}[c]{0.32\textwidth}
                \centering\includegraphics[width=\linewidth]{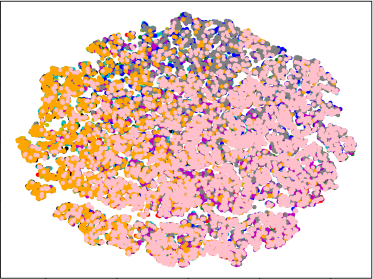}
                \caption*{(e)}
            \end{minipage}\hfill
            \begin{minipage}[c]{0.32\textwidth}
                \centering\includegraphics[width=\linewidth]{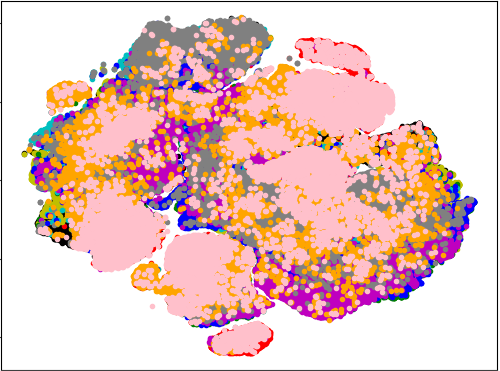}
                \caption*{(f)}
            \end{minipage}\hfill
            \begin{minipage}[c]{0.32\textwidth}
                \centering\includegraphics[width=\linewidth]{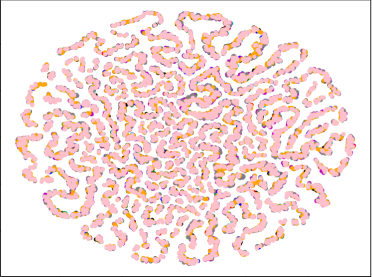}
                \caption*{(g)}
            \end{minipage}
        \end{minipage}%
    }
    \caption{t-SNE visualization of the malicious party’s bottom model output on CIFAR-10. Each color denotes a true label. Clearer separation indicates higher label leakage. (a)~Original model; (b)~GC; (c)~NG; (d)~MG; (e)~FLSG; (f)~KDK; (g)~LADSG; }
    \label{fig:tsne-viz}
\end{figure}

To understand how different defenses reshape the semantic representation space, we visualize the local model outputs using t-SNE on CIFAR-10 (Figure~\ref{fig:tsne-viz}). The results reveal that \textbf{LADSG is the only method that effectively eliminates class-level separability in embeddings}, which aligns with its superior performance in reducing label inference success (Figure~\ref{fig:attack_comparison}).

Without defense, the embeddings form clear “petal-shaped” clusters, indicating strong retention of label semantics. Traditional methods like GC and NG introduce perturbations at the distribution periphery but leave cluster centers intact, suggesting limited disruption to class geometry. FLSG and KDK show stronger effects: surrogate gradients and label obfuscation blur class boundaries and reduce density, but some clusters-especially for semantically distinct classes-remain visible, limiting their overall defense strength.

In contrast, LADSG generates a vortex-like, isotropic embedding where class samples are highly entangled and boundaries are indistinguishable. This structural collapse of class separability reflects the joint effects of gradient and label-path obfuscation, and directly explains the substantial drop in attack success rate. These findings support the necessity of \emph{multi-path semantic perturbation} for robust label protection in VFL.

\subsection{Model Performance and Runtime Efficiency}
\label{Model Performance and Runtime Efficiency}

\begin{figure}[t]
    \centering
    \includegraphics[width=\textwidth]{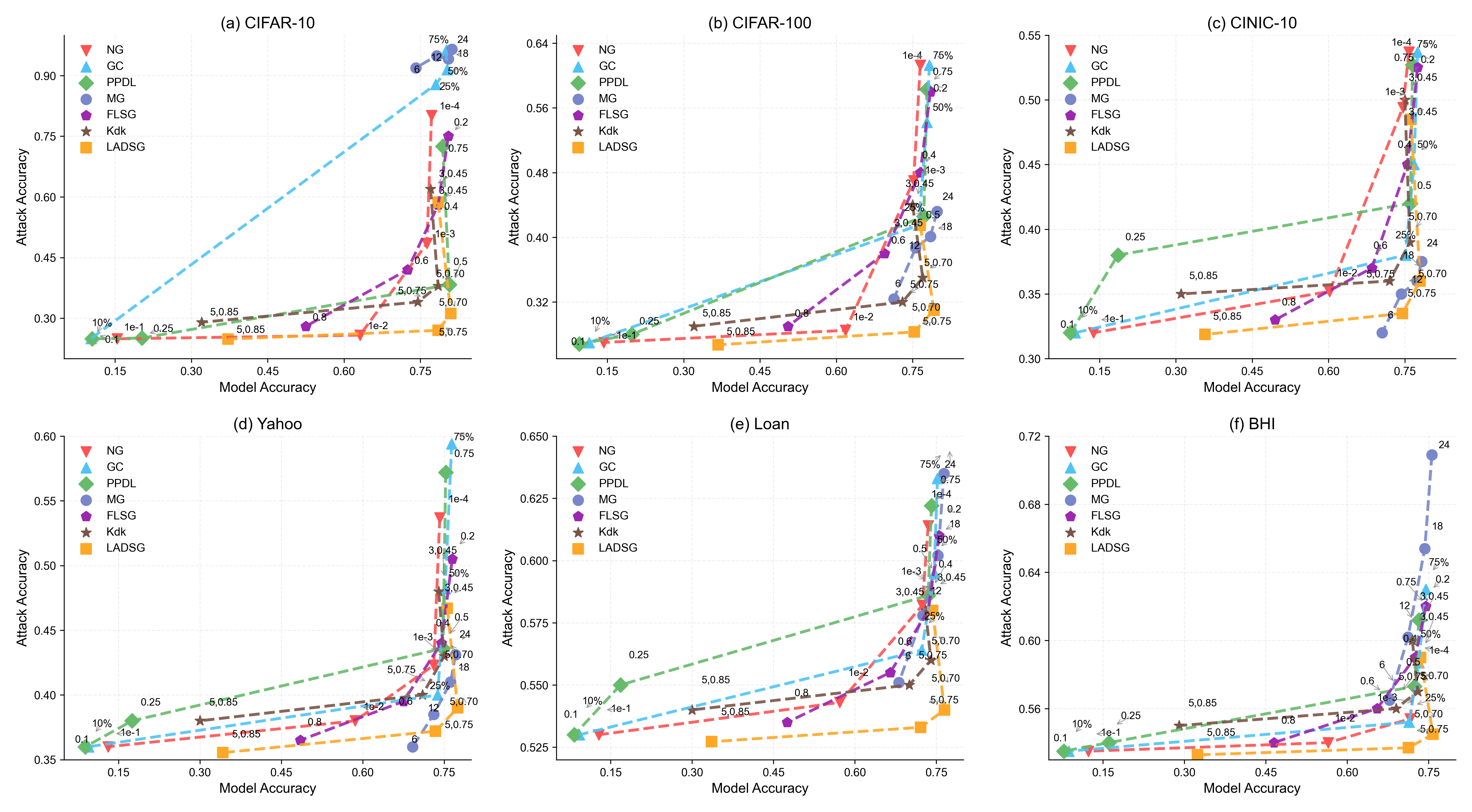}
    \caption{Security-utility trade-off of different defense methods across six datasets. Each curve represents a defense method under varying parameters. Ideal defenses appear in the lower-right (high accuracy, low attack success).}
    \label{fig:model_perf}
\end{figure}

To evaluate the trade-off between label protection and model utility, we conducted systematic experiments across six datasets, tuning each defense mechanism over its representative parameter range. Figure~\ref{fig:model_perf} summarizes the privacy–utility performance of all methods. We observe that \textbf{LADSG consistently achieves the best balance}, maintaining high model accuracy while effectively reducing label inference risks. This is largely attributed to its dual-path obfuscation and lightweight design, which ensure both convergence speed and minimal performance loss.

Traditional methods like GC and NG rely on strong perturbations such as aggressive gradient clipping or independent noise injection. While they reduce leakage to some extent, they also disrupt the gradient propagation process, leading to slower convergence and notable accuracy degradation. For instance, GC’s severe truncation of gradient magnitudes hampers training signal preservation, whereas NG’s unstructured noise is often suppressed in aggregation. More recent defenses, notably FLSG and KDK, pursue higher utility yet still tackle only one leakage channel. FLSG substitutes real gradients with surrogates constrained solely by cosine similarity; this single angular constraint permits sizeable magnitude deviations, slowing convergence. KDK obscures supervision via soft labels, effectively curbing model-side leakage, but-by disregarding gradient exposure-remains susceptible to gradient-based inference.

In contrast, LADSG fuses dual-similarity–constrained surrogate gradients with label-anonymized soft distillation, jointly shielding both gradient- and model-side leakage. This design maintains learning directionality, curtails label semantics, and accelerates convergence. Empirically, LADSG delivers consistent protection and accuracy over a wide range of hyper-parameter choices (e.g., $\epsilon$, $k$, $\tau$), making it a practical, deployment-ready defense.

\begin{figure}[t]
    \centering
    \includegraphics[width=0.65\textwidth]{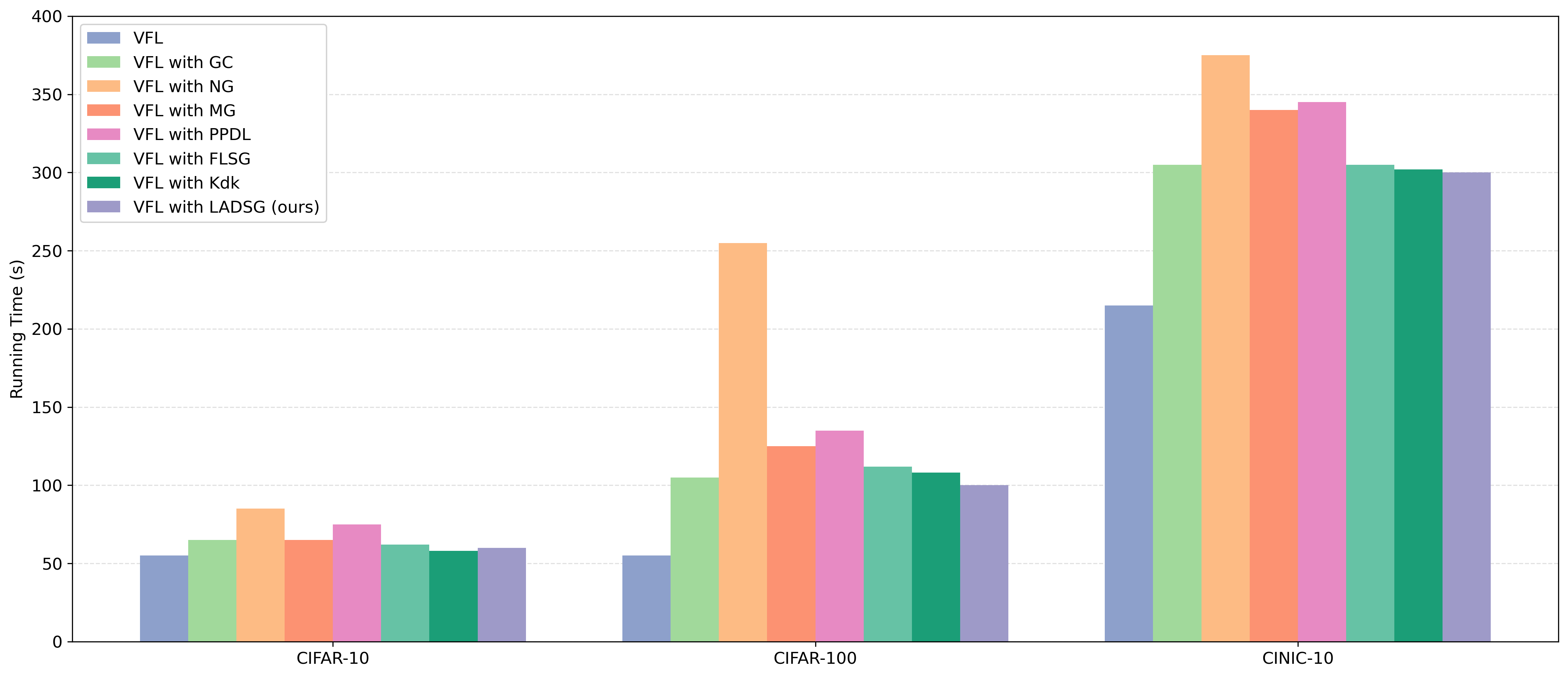}
    \caption{Running time (seconds per epoch) of different defense methods on three datasets. Lower values indicate better efficiency.}
    \label{fig:runtime_comparison}
\end{figure}

To further assess the training efficiency of different defense strategies, we measure the per-round training time on CIFAR-10, CIFAR-100, and CINIC-10, as shown in Figure~\ref{fig:runtime_comparison}. While all defense methods inevitably introduce additional overhead, LADSG consistently demonstrates competitive runtime across all datasets. Notably, on CIFAR-100 and CINIC-10, it significantly outperforms more complex mechanisms such as PPDL and NG, and achieves comparable or even better efficiency than FLSG and KDK. These results indicate that LADSG strikes a favorable balance between defense effectiveness and computational efficiency, highlighting its potential for scalable real-world deployment.

\subsection{Ablation Study: Component Effectiveness}
\label{Ablation Study}

\begin{figure}[t]
    \centering
    \includegraphics[width=\textwidth]{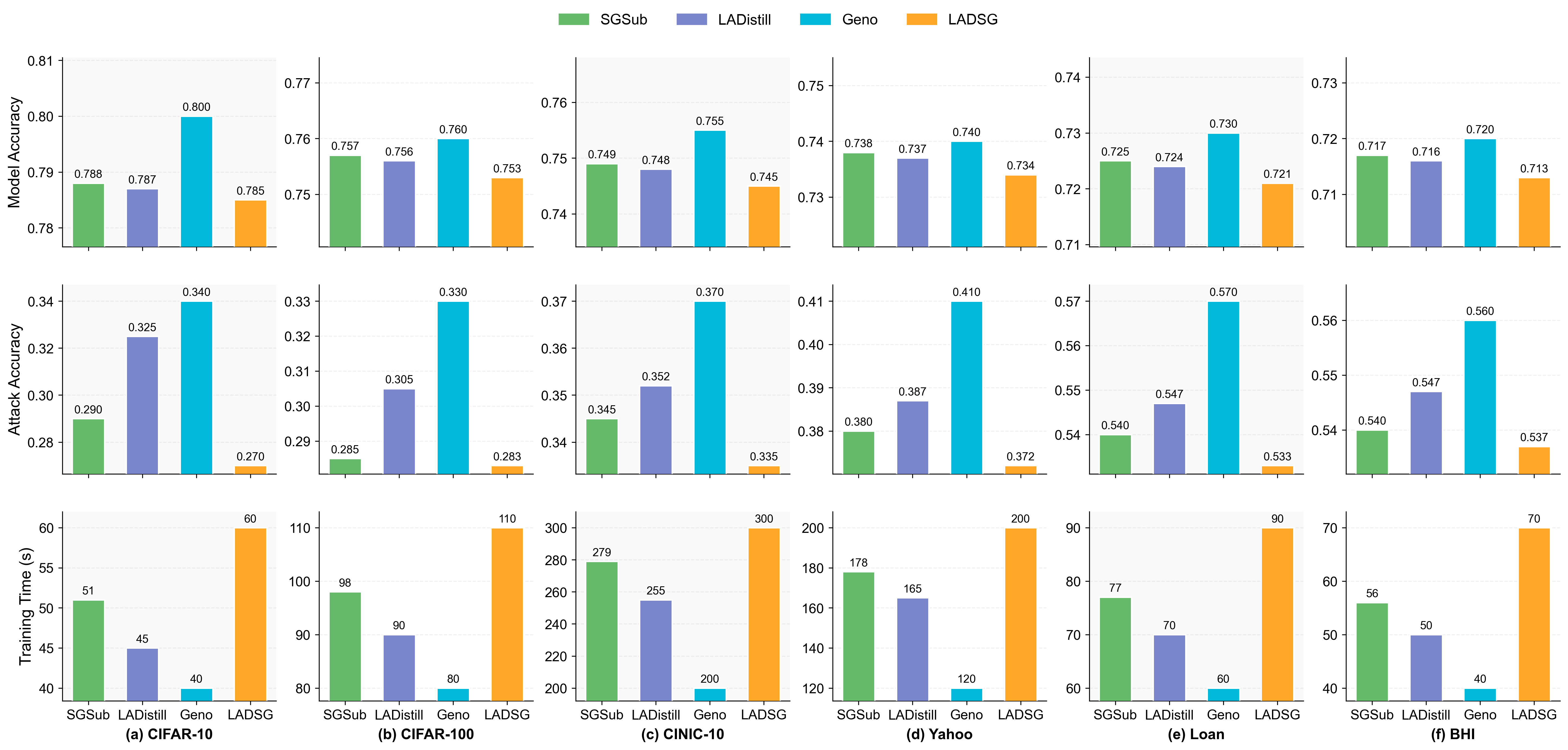}
    \caption{Ablation study on six datasets. Top: model accuracy; middle: attack accuracy; bottom: training time (seconds).}
    \label{fig:ablation_study}
\end{figure}

To assess the effectiveness of each component in our LADSG framework, we perform an ablation study across six datasets. As shown in Figure~\ref{fig:ablation_study}, all three modules-SGSub (gradient substitution), LADistill (label anonymization), and GENO (anomaly detection)-individually contribute to reducing attack success rates without compromising model accuracy. However, only their full integration in LADSG consistently achieves the best trade-off, with the lowest attack success rates and competitive runtime. For example, on CIFAR-10, LADSG reduces the attack success rate to 0.270, compared to 0.290 with SGSub and 0.325 with LADistill alone. This validates the synergistic effect of combining semantic obfuscation and gradient perturbation.

Individually, SGSub effectively disrupts gradient-based leakage by substituting gradients with statistically aligned surrogates, achieving notable semantic confusion. LADistill lowers model-path leakage by replacing true labels with soft distributions, yet lacks influence on backpropagation signals. GENO improves model stability by filtering anomalous gradients, occasionally improving accuracy, but does not sufficiently hinder class-level separability. These limitations explain their partial effectiveness when used in isolation. The integrated LADSG framework achieves substantially better performance than any single component, offering the lowest attack success rates and improved model accuracy, with acceptable training overhead. This demonstrates that only their combination can deliver both robust privacy protection and strong model utility.

\begin{figure}[t]
    \centering
    \begin{subfigure}[t]{0.24\textwidth}
        \centering
        \includegraphics[width=\textwidth]{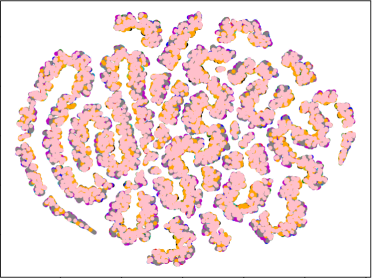}
        \caption{SGSub}
    \end{subfigure}
    \hfill
    \begin{subfigure}[t]{0.24\textwidth}
        \centering
        \includegraphics[width=\textwidth]{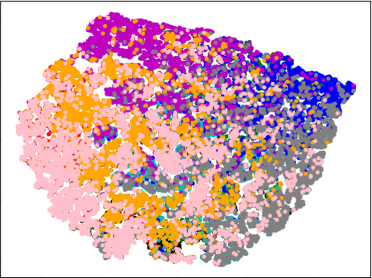}
        \caption{LADistill}
    \end{subfigure}
    \hfill
    \begin{subfigure}[t]{0.24\textwidth}
        \centering
        \includegraphics[width=\textwidth]{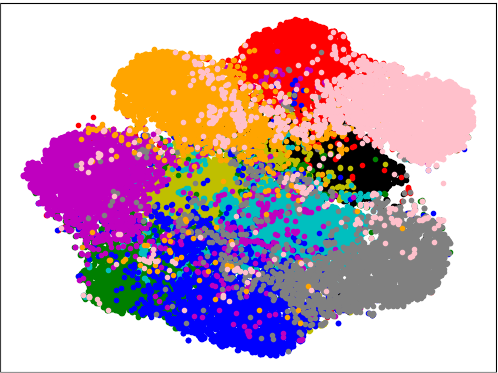}
        \caption{GENO}
    \end{subfigure}
    \hfill
    \begin{subfigure}[t]{0.24\textwidth}
        \centering
        \includegraphics[width=\textwidth]{imgs/our.png}
        \caption{LADSG}
    \end{subfigure}
    \caption{t-SNE visualization of learned representations on CIFAR-10 under ablation settings.}
    \label{fig:tsne_ablation}
\end{figure}

To further illustrate the representation-level effects, we visualize model outputs on CIFAR-10 using t-SNE under different module combinations (Figure~\ref{fig:tsne_ablation}). SGSub leads to entangled spiral-like structures, showing strong semantic disruption. LADistill retains visible clusters due to intact gradient paths. GENO improves stability but sharpens cluster boundaries. Only the full LADSG system produces isotropic, blurred embeddings where class separation is largely erased. This visual evidence corroborates our quantitative findings and highlights the necessity of component interplay in weakening semantic distinguishability.

\section{Conclusion}

This paper presents a systematic study of label leakage in vertical federated learning (VFL), encompassing three major categories of attacks: passive, active, and direct inference. To address this challenge, we propose a novel defense framework, LADSG, which targets the two core pathways of label leakage. LADSG integrates three key modules-semantically aligned fake gradient generation, lightweight anonymous distillation, and abnormal gradient detection-to achieve joint protection against diverse attack types.

Extensive experiments on six real-world datasets demonstrate that LADSG achieves an outstanding balance between privacy protection and model utility. Its integrated tri-module design significantly reduces attack success rates-by up to 60\%-while incurring less than 5\% accuracy loss. Compared to existing defenses, LADSG delivers superior robustness and generalizability across diverse attack scenarios. The modular design provides complementary protection across multiple leakage paths, as confirmed by ablation studies, highlighting its practicality for real-world VFL deployment.

While LADSG provides strong protection against label inference, its current time complexity remains on par with existing methods and could be further optimized. Future work will focus on improving efficiency, deepening theoretical understanding of label leakage, and extending the framework to more complex VFL scenarios such as heterogeneous features and multi-party collaboration.
%
%
%
\bibliographystyle{splncs04}
\bibliography{references}
%
%
%
%
%
\end{document}